# Learning goals and perceived irrelevance to major within life science majors in introductory physics

Andrew J. Mason

*Department of Physics and Astronomy, University of Central Arkansas,*
*201 S. Donaghey Avenue, Conway, AR 72035*

**Abstract:** Previously, students' self-expressed learning goals, defined in the context of a problem solving self-diagnosis exercise that served as a pre-lab activity, were studied as a potential variable that might be related to various course measurements (overall grade, FCI and CLASS pre-post scores) of an introductory algebra-based physics course population primarily consisting of life science majors. In this study, the same student population (218 total students) was polled for students' opinion about what aspects of the course pertained to their majors. Approximately 23% of the course population (50 students) explicitly stated a belief that the course had little or nothing to do with their majors; the other students named a specific physics topic, overall usefulness, or an aspect of the course closely related to well-known PER topics (e.g., problem solving or conceptual understanding), etc. We investigate the belief that the course is irrelevant to one's major as another potential mindset variable, alongside task-specific learning orientations, that influences pre-post course measurements.

## I. INTRODUCTION

The notion of relevance of a university-level course to a student [1] applies to non-physics majors' perceptions of how a physics course is relevant to their major, e.g., an introductory algebra-based physics course sequence whose population is predominately life-science. Recent efforts to transform introductory physics for life sciences (IPLS) courses have included discussing relevance via context of physics topics to specific life science populations, e.g., for biology majors [2] or for health science majors (e.g., physical therapy pre-professionals) [3]. However, not all IPLS populations may be treated as monolithic specific majors, e.g., a course section that has a roughly equal mixture of biology majors and health science majors, which may occur in an institution type that differs in size and/or scope from the large research universities typically featured in IPLS publications.

The topic of student mindset, primarily with regard to student beliefs about intelligence, has a very strong presence in psychology literature [4,5], and may contribute to the conversation of perceived relevance to students. Aspects of student mindset which treat relevance have recently become an explicit topic of interest by PER literature, e.g., student responses about challenge [6], establishing a STEM "identity" [7], or learning and achievement goals [8]. The latter item has been a concern with regard to recognizing achievement goals in the form of learning orientations, which were previously indicated to be a factor in the success of upper-level graduate students in physical science programs [9], but not necessarily at the introductory level [8].

Learning goal orientations for life science majors in an algebra-based introductory physics course were identified via surveying students' perception of the utility of a pre-lab problem solving exercise [8]. This exercise was based upon research in "self-diagnosis" metacognition in physics [10, 11] with the use of context-rich problems [12] and the assistance of the instructor and lab partners. Students were asked to solve context-rich problems during the first portion of their lab periods, then use a scoring rubric to determine their strengths and weaknesses during the problem-solving attempt, prior to carrying out a lab exercise which was conceptually linked to the pre-lab exercise. The results in Ref. 8 indicated that learning goals were identifiable from students' short-answer written responses to a feedback survey question on the problem solving exercise, and moreover that they appeared to be validated in terms of representing differences in attitudinal gains on problem-solving item clusters within the CLASS survey [13], as well as on conceptual understanding clusters and over the entire survey. However, this correlation did not translate to any perceivable effect on pre-post measurements for the FCI [14] or on overall course performance.

### A. Current Research Goals

In this follow-up study, we supplement the study in Ref. 8 by examining another question given to students on the same post-test feedback survey: "Which sections of the [course] material did you find related well to the coursework in your major?" A non-trivial percentage of student respondents expressed that the course in fact had nothing to do with their majors; they perceived the course as irrelevant. This course section provides an excellent opportunity to consider relevance in a straightforward fashion – namely, whether students who have already committed to a STEM discipline for a major see any relevance of physics to their own choice of major.

Therefore, our first research question is to determine how many biology majors and health science majors, respectively, viewed an introductory physics course as irrelevant to their course major, and to see if this factor affects pre-post measurements within each choice of major. Our second question is to similarly consider whether each subgroup of learning orientations identified in Ref. 8 had a significant influence of student perception of irrelevance, and to what extent these subgroups of students may affect statistical results.

## II. PROCEDURE

The sample consisted of 218 students, distributed over six semesters' worth of introductory first-semester algebra-based physics course sections, all with the same course instructor and lecture/laboratory course format. The criteria for inclusion of data from students was that students must a) complete and pass the course, b) submit completed pre-post data (described below) as part of required coursework, and c) not observably give unserious answers on any survey data (usually per a CLASS item that exists for that purpose). Students were given full points for participation, so as to avoid making their responses grade-dependent.

During the last laboratory session of the semester for an introductory algebra-based physics course at a primarily undergraduate institution, students submitted an in-class written response to the following post-test survey question regarding the aforementioned pre-lab metacognitive problem solving exercise: "In what ways did you find this exercise useful towards learning the material in the course?" Three learning goal orientation groups were determined by inter-rater reliability of student responses to this question (see Ref. 8 for details): interest in mastery of a problem solving framework per se, i.e., "Framework" orientation; interest in doing well in terms of course points, i.e.,

"Performance" orientation; and students who did not express a learning goal, but focused on collaborative process or other items, i.e., "Vague" orientation.

Students also answered a follow-up question: "Which sections of the [class] material did you find related well to the coursework in your major?" In this data set, 50 out of 218 students, or about 23% of the course population, declared that nothing within the course related to their majors, thus expressing a perception of irrelevance; reasons for this statement included, but were not limited to, statements that one's major wasn't really a science major to begin with (e.g., describing one's health administration major as being primarily a business major), or stating that one's science major wasn't really involved with physics (e.g., a pre-veterinarian biology major not perceiving relevance of physics towards animal health). The remaining students (168 total) cited either a specific physics topic, a specific skill emphasized by the course (e.g., problem solving, conceptual topics, real-world applications), or a specific application to their majors from the course as being relevant to their majors (e.g., torque as applied to physical therapy of limbs).

The CLASS survey [13] and FCI [14] were administered in pre-post format on the first and last weekly laboratory sessions of the semester, respectively. In this paper, we will examine overall FCI performance, overall CLASS performance, and analysis of specific item clusters on the CLASS (as certain item clusters, e.g., Personal Interest and Real-World Context, consider the perception of relevance of physics to the respondent's own life experience).

## III. RESULTS

### A. Distribution of Perceptions of Relevance vs. Irrelevance

Table 1 shows the distributions of students who saw the physics course as relevant, and students who saw the course as irrelevant. The distributions are shown in terms of the overall student pool; in terms of the subgroups based upon choice of major; and in terms of "Framework," "Performance," and "Vague" learning orientations of students towards the pre-lab problem-solving exercise [8]. Biology majors were slightly less likely to perceive irrelevance than were health science majors, and Vague-oriented students were slightly more likely to perceive irrelevance than were other learning orientations. However, there is no statistical separation between groups for either majors or orientations.

TABLE 1. Distributions of students who saw the physics course as relevant in some way to their majors vs. students who saw the course as irrelevant. Students who perceived irrelevance in each respective group are displayed both in terms of raw number and in terms of the percentage out of the entire group.

| Group (n) | Relevant | Irrelevant | Irr. % |
|---|---|---|---|
| All (218) | 168 | 50 | 22.9% |
| Biology (91) | 76 | 15 | 16.5% |
| Health (85) | 66 | 19 | 22.4% |
| Other Sci (35) | 21 | 14 | 40.0% |
| Non-Sci (7) | 5 | 2 | 28.6% |
| Framework (76) | 59 | 17 | 22.4% |
| Performance (79) | 63 | 16 | 20.3% |
| Vague (63) | 46 | 17 | 27.0% |

### B. Effect of Relevance on FCI Performance

Table 2 examines the FCI pre-test, post-test, and normalized gain average for the "Relevant" population versus the "Irrelevant" population. A check of t-tests between groups, as well as a Cohen's d calculation for effect size, are presented. There are no statistically significant differences on the FCI between pretest scores (which are close to random-chance), post-test scores, or normalized gains.

An examination of relevance vs. irrelevance within biology and health science majors did not show any statistical differences in force concept understanding between the two sets of life science majors. A similar examination for relevance vs. irrelevance within learning goal orientations (Framework, Performance, and Vague) showed no statistical significance on pretest scores, posttest scores, or normalized gains for any orientation. A moderate effect size ($d \sim 0.3$) appeared to show a relatively better score for relevant-perceiving students than for irrelevance-perceiving students within the Framework-oriented and Vague-oriented groups.

### C. Effect of Relevance on CLASS Performance

Table 3 examines the CLASS pre-test, post-test, and normalized gain average for the "Relevant" population versus the "Irrelevant" population, in terms of the overall survey results and in terms of each item cluster within the CLASS. We include $t$-test comparisons for statistical significance and Cohen's $d$ value for effect size. All comparisons featured were determined to have statistically equal variances via $F$-test.

We find that perception of relevance does not appear to be a statistically significant factor on the pretest,

TABLE 2. Average FCI pretest and posttest score percentages and normalized gains, comparing between students with a "Relevant" view of physics and students with an "Irrelevant" view. All comparisons had equal variance ($F$-test: $p > 0.05$).

| FCI result | Irrelevant ($n = 50$) | Relevant ($n = 168$) | Cohen's $d$ ($p$-value) |
|---|---|---|---|
| Pretest | 24.5% | 26.2% | 0.12 (0.44) |
| Posttest | 36.0% | 37.8% | 0.11 (0.49) |
| Gain (g) | +14.5 | +15.5 | 0.06 (0.70) |

TABLE 3. Average CLASS pre-test and post-test percentages of expert-like scores and normalized gains, comparing between students who expressed "Relevance" and students who expressed "Irrelevance" in response to the post-test survey question "Which sections of the [class] material did you find related well to the coursework in your major?" The abbreviations for the item clusters are explained in Ref. 13. A positive gain indicates a net expert-like shift, and a negative gain indicates a net novice-like shift.

| Pretest scores ($n$) | Overall | PI | RWC | PS-G | PS-C | PS-S | SME | CU | ACU |
|---|---|---|---|---|---|---|---|---|---|
| Irrelevance % (50) | 54.3 | 44.0 | 64.0 | 58.8 | 58.5 | 39.7 | 65.7 | 51.0 | 37.4 |
| Relevance % (168) | 57.6 | 51.6 | 65.5 | 63.9 | 65.6 | 43.7 | 70.5 | 54.9 | 43.1 |
| Effect size | 0.22 | *0.29* | 0.05 | 0.23 | 0.26 | 0.15 | 0.22 | 0.15 | 0.25 |
| $p$-value | 0.18 | *0.08* | 0.76 | 0.16 | 0.11 | 0.34 | 0.18 | 0.36 | 0.13 |
| **Posttest scores ($n$)** | **Overall** | **PI** | **RWC** | **PS-G** | **PS-C** | **PS-S** | **SME** | **CU** | **ACU** |
| Irrelevance % (50) | 50.0 | 32.7 | 50.0 | 51.5 | 51.0 | 32.7 | 60.0 | 47.0 | 36.0 |
| Relevance % (168) | 58.1 | 47.8 | 63.2 | 61.5 | 62.9 | 42.5 | 68.1 | 56.5 | 43.5 |
| Effect size | **0.53** | **0.52** | *0.40* | *0.41* | *0.39* | *0.37* | *0.33* | *0.36* | *0.31* |
| $p$-value | **<0.01** | **<0.01** | **0.01** | **0.01** | **0.02** | **0.02** | **0.04** | **0.03** | *0.05* |
| **Normalized gains ($n$)** | **Overall** | **PI** | **RWC** | **PS-G** | **PS-C** | **PS-S** | **SME** | **CU** | **ACU** |
| Irrelevance % (50) | -0.06 | -0.24 | -0.15 | -0.04 | -0.03 | -0.21 | -0.03 | -0.10 | -0.14 |
| Relevance % (168) | +0.04 | -0.08 | +0.06 | +0.03 | +0.02 | -0.07 | +0.08 | +0.06 | -0.04 |
| Effect size | **0.37** | **0.33** | **0.36** | 0.14 | 0.10 | *0.28* | 0.24 | **0.33** | 0.22 |
| $p$-value | **0.02** | **0.04** | **0.03** | 0.38 | 0.56 | *0.08* | 0.14 | **0.04** | 0.18 |

either overall or for any item cluster. On the other hand, there is a significant difference on the posttest, both overall and within almost every item cluster, with moderate effect sizes according to Cohen's *d* calculations, and in all cases with a more expert-like view of physics expressed by the students who perceived "Relevance" than by the students who perceived "Irrelevance."

There is also a statistical significance in normalized gains, as well as a moderate effect size difference, for the overall CLASS. Students who perceived "Relevance" to the course experienced a slight expert-like shift, while students who perceived "Irrelevance" expressed a slight novice-like shift. This pattern of a more novice-like shift for an "Irrelevance" perception appeared to be somewhat demonstrated across all item clusters, although not always with statistical significance. In particular, the three item clusters that demonstrated the strongest differences between "relevant" and "irrelevant" beliefs were Personal Interest (PI), Real-World Connection (RWC), and Conceptual Understanding (CU). The first two clusters are of significance because they do not necessarily reflect the pre-lab problem solving exercise about which the students were specifically asked; they are more directly related to explicit relevance concerns within student responses. Even so, the effect sizes for these findings are only moderate ($d \sim 0.33$ in each case).

For a closer examination, we consider the CLASS results within each choice of life science major and within each learning orientation. Table 4 shows the overall CLASS results for choice of life science major, and Table 5 shows overall CLASS results for learning orientations.

Table 4 indicates that each life science major separately exhibits the pattern within the overall CLASS results in Table 3, between relevant-perceiving and irrelevant-perceiving students. Both sets of majors show no statistical significance on the pretest, but a large effect size and significance on the posttest (with more expert-like scores from relevant-perceiving students in both cases).

There is also a large effect size and statistical significance for normalized gains within the biology majors, again with a novice-like shift for irrelevant-perceiving students and a slight expert-like shift for relevant-perceiving students. This pattern also appears for health science majors' gains, but is not statistically significant in this case. With regard to CLASS item clusters, biology majors who perceived relevance had statistically higher post-test scores and normalized gains on the PI and RWC clusters, as well as on the Sense Making-Effort (SME) cluster, with large effect sizes for these three clusters ($0.6 < d < 0.85$ for post-test comparisons, $0.5 < d < 0.8$ for gains comparisons).

TABLE 4. Average CLASS pretest and posttest percentages of expert-like scores and normalized gains, comparing students who perceived relevance vs. students who did not.

| | Group | Irrelevant ($n$) | Relevant ($n$) |
|---|---|---|---|
| Pretest (%) | Biology | 55.7 (16) | 61.1 (75) |
| | Health | 48.1 (19) | 52.3 (66) |
| Posttest (%) | Biology | **49.0 (16)** | **61.7 (75)** |
| | Health | **43.4 (19)** | **52.5 (66)** |
| Gain ($g$) | Biology | **-.11 (16)** | **+.06 (75)** |
| | Health | -.09 (19) | +.02 (66) |
| Cohen's *d* effect sizes ($p$-values from *t*-tests) | | | |
| | Biology | | Health |
| Pretest | 0.39 (0.16) | | 0.29 (0.27) |
| Posttest | **0.79 (< 0.01)** | | **0.52 (< 0.05)** |
| Gains | **0.58 (< 0.05)** | | 0.39 (0.14) |

TABLE 5. Average CLASS pretest and posttest percentages of expert-like scores and normalized gains, comparing students who perceived relevance vs. students who did not.

| | Group | Irrelevant (n) | Relevant (n) |
|---|---|---|---|
| Pretest (%) | Framework | 53.4 (17) | 58.1 (59) |
| | Performance | 57.4 (16) | 57.1 (63) |
| | Vague | 52.1 (17) | 57.6 (46) |
| Posttest (%) | Framework | *54.9 (17)* | *62.6 (59)* |
| | Performance | *46.5 (16)* | *56.1 (63)* |
| | Vague | 48.2 (17) | 54.9 (46) |
| Gains (g) | Framework | +.05 (17) | +.13 (59) |
| | Performance | **-.18 (16)** | **+.01 (63)** |
| | Vague | -.06 (17) | -.02 (46) |
| **Cohen's *d* effect sizes (*p*-values from *t*-tests)** | | | |
| | Framework | Performance | Vague |
| Pretest | 0.35 (0.21) | 0.02 (0.94) | 0.35 (0.22) |
| Posttest | *0.47 (0.10)* | *0.53 (0.07)* | 0.40 (0.17) |
| Gains | 0.26 (0.35) | **0.68 (0.02)** | 0.18 (0.54) |

Table 5 shows that the difference between relevant and irrelevant perceptions appears to particularly matter for Performance-oriented students' attitudinal shifts (presented in terms of normalized gains on the percentage of expert-like responses); students who perceive relevance experience little change on average, but students who perceive irrelevance have a rather noticeable novice-like shift difference (with statistical significance and a large effect size). There was no similar effect within framework-oriented students (who had a slight expert-like gain in either case) or within vague-oriented students (who had a slight novice-like gain in either case). Within the item clusters, the Performance-oriented students had statistical significance ($p < 0.05$) or borderline significance ($0.05 < p < 0.1$) across all item clusters, with students who perceived irrelevance having a more novice-like shift in each case.

## IV. DISCUSSION

While most students do perceive at least some relevance within a physics course towards their respective majors, a non-trivial percentage of students believe the course is irrelevant. Students who perceive irrelevance can have a significant effect on evaluating CLASS results from student populations. Relevance vs. irrelevance perception appears to have a moderate, statistically significant effect on attitudinal shifts for the first-semester algebra-based introductory physics course population that is predominately biology and health science majors. The robust effect appears to be sustained when looking at biology majors and (to a somewhat lesser extent) for health science majors.

The issue of relevance also appears to be important as a co-factor alongside learning goal orientation within the CLASS data. Students who are performance-oriented in particular appear more likely to have a novice-like shift if they perceive that the course is irrelevant to their choice of major. Students who are mastery-oriented (i.e., Framework-oriented in this context) do not appear to rely upon perceiving relevance, as they appear genuinely interested in mastering the material and appear to have an expert-like shift on average regardless. Conversely, a lack of a learning goal (i.e., Vague-oriented students) also seems to not depend on perception of relevance.

The lack of differentiation between students in this population on FCI pre-post gains, as observed in Ref. 8, persists with regard to students' perceived relevance vs. irrelevance of physics with respect to their majors. It appears that neither learning goals nor relevance is a factor as far as Force Concept Inventory responses indicate within the scope of a single physics course.

This follow-up study has three limitations to consider for future research and implementations. First, the study is done in a primarily undergraduate institution, and the sampled course is taught by an undergraduate department. Physics departments in similar circumstances will benefit in recognizing whether or not biology majors and health science majors must have their own respective course sequences designed for them, as well as which specific majors (e.g., health administration, which is primarily a business major at the host institution) in fact require algebra-based physics generally speaking. Different institution types may also determine whether learning orientations and relevance are a factor in the effectiveness of pedagogical course implementations, e.g., with large IPLS courses that are monolithically populated by a single life science major. Second, the measurements of relevance and learning orientation were determined with post-test data to examine where students were at the end of the course, with learning goals orientation determined from student responses to a specific pre-lab exercise that was based on research in metacognitive problem solving techniques. This suggests a more general identification of learning goal orientations is necessary, beyond the scope of a specific pedagogical implementation. Third, a pretest measurement of learning goals and relevance may be necessary to explore whether they are static or growth-related aspects of mindset [4,7], as student mindset may be shaped over the course of a semester if perceived relevance or irrelevance changes.

## ACKNOWLEDGEMENTS

The author thanks the UCA Department of Physics and Astronomy for funding support, and Charles A. Bertram for his pertinent contributions to Ref. 8.

[1] M. Stuckey, A. Hofstein, R. Mamlok-Naaman, and I. Eilks, "The meaning of 'relevance' in science education and its implications for the science curriculum." Studies in Sci. Ed. **49** (1), 1-34 (2013).

[2] e.g., E.F. Redish et al., "NEXUS/Physics: An interdisciplinary repurposing of physics for biologists," Am. J. Phys. **82**, 368 (2014); C.H. Crouch and K. Heller, "Introductory physics in biological context: An approach to improve introductory physics for life science students," Am. J. Phys. **82**, 378 (2014).

[3] E. Mylott, E. Kutschera, J. Dunlap, W. Christensen, and R. Widenhorn, "Using biomedically relevant multimedia content in an introductory physics course for life science and pre-health students," J. Sci. Ed. Tech. **25**(2), 222 (2016).

[4] C. Dweck, "Mindset: The New Psychology of Success." Ballantine, New York (2007).

[5] D. Belenky and T. Nokes-Malach, "Motivation and transfer: The role of mastery-approach goals in preparation for future learning." J. Learn. Sci. **21**, 399 (2012).

[6] A. Little, B. Humphrey, A. Green, A. Nair, and V. Sawtelle, "Exploring mindset's applicability to students' experiences with challenge in transformed college physics courses." Phys. Rev. PER **15**, 010127 (2019).

[7] D. Bennett, L. Roberts, and C. Creagh, "Exploring possible selves in a first-year physics foundation class: Engaging students by establishing relevance." Phys. Rev. PER **12**, 010120 (2016).

[8] A. Mason and C. Bertram, "Consideration of learning orientations as an application of achievement goals in evaluating life science majors in introductory physics." Phys. Rev. PER **14**, 010125 (2018). Also see A. J. Mason, Potential Relationship of Chosen Major to Problem Solving Attitudes and Course Performance, 2015 PERC Proceedings [College Park, MD, July 29-30, 2015], edited by A. D. Churukian, D. L. Jones, and L. Ding, doi:10.1119/perc.2015.pr.049; A. J. Mason and C. A. Bertram, Potential relationship of epistemic games to group dynamics and learning orientations towards physics problem solving, 2016 PERC Proceedings [Sacramento, CA, July 20-21, 2016], edited by D. L. Jones, L. Ding, and A. Traxler, doi:10.1119/perc.2016.pr.051; C. A. Bertram and A. J. Mason, The Effect of Students' Learning Orientations on Performance in Problem Solving Pedagogical Implementations, 2017 PERC Proceedings [Cincinnati, OH, July 26-27, 2017], edited by L. Ding, A. Traxler, and Y. Cao, doi:10.1119/perc.2017.pr.009.

[9] Z. Hazari, G. Potvin, R. Tai, and J. Almarode, "For the love of learning science: Connecting learning orientation and career productivity in physics and chemistry." Phys.Rev. ST - Phys. Educ. Res. **6**, 010107 (2010).

[10] E. Yerushalmi, E. Cohen, A. Mason, and C. Singh, "What do students do when asked to diagnose their mistakes? Does it help them? I. An atypical quiz context." Phys. Rev. PER **8**, 020109 (2012).

[11] E. Yerushalmi, E. Cohen, A. Mason, and C. Singh, "What do students do when asked to diagnose their mistakes? Does it help them? II. A more typical quiz context." Phys. Rev. PER **8**, 020110 (2012).

[12] P. Heller and M. Hollabaugh, "Teaching problem solving through cooperative grouping. Part 2: Designing problems and structuring groups." Am. J. Phys. **60**, 637 (1992).

[13] W. Adams, K. Perkins, N. Podolefsky, M. Dubson, N. Finkelstein, and C. Wieman, "New instrument for measuring student beliefs about physics and learning physics: The Colorado Learning Attitudes about Science Survey." Phys. Rev. ST Phys. Educ. Res. 2, 010101 (2006).

[14] D. Hestenes, M. Wells, and G. Swackhamer, "Force concept inventory." Phys. Teach. **30**, 141 (1992).